\documentclass[conference]{IEEEtran}
\IEEEoverridecommandlockouts

\usepackage{float}
\usepackage{cite}
\usepackage{amsmath,amssymb,amsfonts}
\usepackage{graphicx}
\usepackage{textcomp}
\usepackage{xcolor}
\usepackage{amsmath,amssymb,amsfonts}
\usepackage{graphicx}
\usepackage{textcomp}
\usepackage{xcolor}
\usepackage{acro}
\usepackage{stfloats}
\usepackage[capitalise]{cleveref}
\usepackage{siunitx}
\usepackage{algorithm}
\usepackage{algpseudocode}
\usepackage{newtxmath}
\usepackage{enumitem}
\usepackage{dsfont}

\DeclareAcronym{MC}{
    short = MC,
    long  = molecular communication
}
\DeclareAcronym{RX}{
    short = RX,
    long  = receiver
}
\DeclareAcronym{TX}{
    short = TX,
    long  = transmitter
}
\DeclareAcronym{DBMC}{
    short = DMC,
    long  = diffusive \ac{MC}
}
\DeclareAcronym{SM}{
    short = SM,
    long  = signaling molecule
}
\DeclareAcronym{PBS}{
    short = PBS,
    long  = particle-based simulation
}
\DeclareAcronym{ODE}{
    short = ODE,
    long  = ordinary differential equation
}
\DeclareAcronym{EM}{
    short = EM,
    long  = electromagnetic
}
\DeclareAcronym{ND}{
    short = ND,
    long  = nanodevice
}
\DeclareAcronym{LED}{
    short = LED,
    long  = light-emitting diode
}
\DeclareAcronym{OOK}{
    short = OOK,
    long  = On-Off keying
}
\DeclareAcronym{BER}{
    short = BER,
    long  = bit error rate
}
\DeclareAcronym{ISI}{
    short = ISI,
    long  = inter-symbol interference
}
\DeclareAcronym{SG}{
    short = SG,
    long  = stochastic geometry
}
\DeclareAcronym{MISO}{
    short = MISO,
    long  = multiple input single output
}
\DeclareAcronym{RLL}{
    short = RLL,
    long  = run-length limited
}
\DeclareAcronym{MAM}{
    short = MAM,
    long  = memory-aware modulation
}
\DeclareAcronym{GI}{
    short = GI,
    long  = guard interval
}
\DeclareAcronym{TMDS}{
    short = TMDS,
    long  = transition-minimized differential signaling
}
\DeclareAcronym{FAR}{
    short = FAR,
    long  = fully-absorbing receiver
}
\DeclareAcronym{MEM}{
    short = MEM,
    long  = memory-erasing modulation
}
\DeclareAcronym{ML}{
    short = ML,
    long  = maximum likelihood
}
\DeclareAcronym{SISO}{
    short = SISO,
    long  = single-input single-output
}
\DeclareAcronym{FDM}{
    short = FDM,
    long  = finite difference method
}
\DeclareAcronym{OOK-GI}{
    short = OOK-GI,
    long = OOK with guard interval (GI)
}

\makeatletter
\g@addto@macro\normalsize{%
  \setlength{\abovedisplayskip}{1pt}%
  \setlength{\belowdisplayskip}{1pt}%
  \setlength{\abovedisplayshortskip}{1pt}%
  \setlength{\belowdisplayshortskip}{1pt}%
}
\makeatother

\makeatletter
\AddToHook{cmd/appendix/before}{\def\cref@section@alias{appendix}\def\cref@subsection@alias{appendix}}
\makeatother

\begin{document}
\title{Modulation Schemes for Functionalized \\ Vesicle-based MC Transmitters}
\author{\IEEEauthorblockN{Teena~tom~Dieck, Lukas~Brand, Sebastian~Lotter, Kathrin~Castiglione, Robert~Schober, and Maximilian~Schäfer\\}
\IEEEauthorblockA{\small Friedrich-Alexander-Universität Erlangen-Nürnberg}}

\maketitle

\newcommand{\Hplus}[0]{\mathrm{H^{+}}}
\newcommand{\Sub}[0]{\mathrm{SM}}
\newcommand{\I}[0]{\mathrm{I}}
\renewcommand{\L}[0]{\mathrm{L}}
\newcommand{\NA}[0]{\mathrm{N}_{\mathrm{A}}}

\newcommand{\Chin}[1]{C^{\mathrm{H^{+}}}_{\mathrm{in}}(#1)}
\newcommand{\ChinNot}[0]{C^{\mathrm{H^{+}}}_{\mathrm{in}}}
\newcommand{\CsinNot}[0]{C^{\mathrm{S}}_{\mathrm{in}}}
\newcommand{\Chout}[1]{C^{\mathrm{H^{+}}}_{\mathrm{out}}(#1)}
\newcommand{\ChoutSVS}[1]{C^{\mathrm{H^{+}}}_{\mathrm{out,SVS}}(#1)}
\newcommand{\ChinZERO}[0]{C^{\mathrm{H^{+}}}_{\mathrm{in, 0}}}
\newcommand{\Csin}[1]{C^{\Sub}_{\mathrm{in}}(#1)}
\newcommand{\CsinZERO}[0]{C^{\Sub}_{\mathrm{in, 0}}}
\newcommand{\Csout}[1]{C^{\Sub}_{\mathrm{out}}(#1)}
\newcommand{\CsoutMean}[1]{\bar{C}^{\Sub}_{\mathrm{out}}(#1)}
\newcommand{\CsoutSVS}[1]{C^{\Sub}_{\mathrm{out,SVS}}(#1)}
\newcommand{\Csouttot}[1]{C^{\Sub}_{\mathrm{out,tot}}(#1)}
\newcommand{\CsoutNot}[0]{C^{\Sub}_{\mathrm{out}}}
\newcommand{\CsoutBar}[1]{\bar{C}^{\Sub}_{\mathrm{out}}(#1)}
\newcommand{\ChoutZERO}[0]{C^{\mathrm{H^{+}}}_{\mathrm{out, 0}}}
\newcommand{\ChinM}[2]{C^{\mathrm{H^{+}}}_{\mathrm{in},#1}(#2)}
\newcommand{\ChoutM}[2]{C^{\mathrm{H^{+}}}_{\mathrm{out},#1}(#2)}
\newcommand{\ChoutTot}[1]{C^{\mathrm{H^{+}}}_{\mathrm{out,tot}}(#1)}
\newcommand{\CsinM}[2]{C^{\Sub}_{\mathrm{in},#1}(#2)}
\newcommand{\CsoutM}[2]{C^{\Sub}_{\mathrm{out},#1}(#2)}
\newcommand{\CsoutQ}[2]{C^{\Sub}_{\mathrm{out,tot},#1}(#2)}

\newcommand{\CZERO}[0]{C_{0}} %

\newcommand{\Ciin}[1]{C^{\mathrm{I}}_{\mathrm{in}}(#1)}
\newcommand{\CiinM}[2]{C^{\mathrm{I}}_{\mathrm{in},#1}(#2)}
\newcommand{\Ciout}[1]{C^{\mathrm{I}}_{\mathrm{out}}(#1)}
\newcommand{\CioutM}[2]{C^{\mathrm{I}}_{\mathrm{out},#1}(#2)}

\newcommand{\ihp}[1]{i^{\mathrm{H^{+}}}_{\mathrm{E}}(#1)}
\newcommand{\ihl}[1]{i^{\mathrm{H^{+}}}_{\mathrm{L}}(#1)}
\newcommand{\ihsym}[1]{i^{\mathrm{H^{+}}}_{\mathrm{R}}(#1)}
\newcommand{\issym}[1]{i^{\Sub}_{\mathrm{R}}(#1)}

\newcommand{\jhla}[0]{j^{a}_{\mathrm{L}}}
\newcommand{\jhlb}[0]{j^{b}_{\mathrm{L}}}
\newcommand{\jhpa}[0]{j^{a}_{\mathrm{P}}}
\newcommand{\jhpb}[0]{j^{b}_{\mathrm{P}}}
\newcommand{\jhsymb}[0]{j^{b}_{\mathrm{Sym}}(t)}
\newcommand{\jhlaTick}[0]{j^{a*}_{\mathrm{L}}}
\newcommand{\jhlbTick}[0]{j^{b*}_{\mathrm{L}}}
\newcommand{\jhpaTick}[0]{j^{a*}_{\mathrm{P}}}
\newcommand{\jhpbTick}[0]{j^{b*}_{\mathrm{P}}}
\newcommand{\jhsymbTick}[0]{j^{b*}_{\mathrm{Sym}}(t)}

\newcommand{\ghp}[0]{\gamma_{\mathrm{P}}}
\newcommand{\ghl}[0]{\gamma_{\mathrm{L}}}
\newcommand{\ghsym}[0]{\gamma^{\mathrm{H^{+}}}_{\mathrm{Sym}}}
\newcommand{\gssym}[0]{\gamma^{\Sub}_{\mathrm{Sym}}}
\newcommand{\gssymT}[1]{\gamma^{\Sub}_{\mathrm{Sym}}(#1)}
\newcommand{\ghsymT}[1]{\gamma^{\mathrm{H^{+}}}_{\mathrm{Sym}}(#1)}

\newcommand{\ghpM}[1]{\gamma_{\mathrm{P},#1}}
\newcommand{\ghlM}[1]{\gamma_{\mathrm{L},#1}}
\newcommand{\ghlHatM}[1]{\hat{\gamma}_{\mathrm{L},#1}}
\newcommand{\ghsymM}[1]{\gamma^{\mathrm{H^{+}}}_{\mathrm{Sym},#1}}
\newcommand{\gssymM}[1]{\gamma^{\Sub}_{\mathrm{Sym},#1}}

\newcommand{\ghpHat}[0]{\hat{\gamma}_{\mathrm{P}}}
\newcommand{\ghlHat}[0]{\hat{\gamma}_{\mathrm{L}}}
\newcommand{\ghlHatMean}[0]{\bar{\hat{\gamma}}_{\mathrm{L}}}
\newcommand{\ghsymHat}[0]{\hat{\gamma}^{\mathrm{H^{+}}}_{\mathrm{Sym}}}
\newcommand{\gssymHat}[0]{\hat{\gamma}^{\Sub}_{\mathrm{Sym}}}

\newcommand{\fSym}[0]{\nu_{\mathrm{Sym}}}

\newcommand{\Nh}[0]{N^{\mathrm{H^{+}}}}
\newcommand{\Ns}[0]{N^{\Sub}}
\newcommand{\NsM}[1]{N^{\Sub}_{#1}}

\newcommand{\p}[1]{p(#1)}
\newcommand{\np}[0]{n_{\mathrm{P}}} %
\newcommand{\npMean}[0]{\bar{n}_{\mathrm{P}}} %
\newcommand{\npM}[1]{n_{\mathrm{P},#1}} %
\newcommand{\ntotM}[1]{n_{\mathrm{tot},#1}} %
\newcommand{\ntot}[0]{n_{\mathrm{tot}}}
\newcommand{\thresh}[0]{\xi}

\newcommand{\Km}[0]{\mathrm{K}_{\mathrm{m}}}
\newcommand{\nsym}[0]{n_{\mathrm{Sym}}}
\newcommand{\nsymMean}[0]{\bar{n}_{\mathrm{Sym}}}
\newcommand{\nsymM}[1]{n_{\mathrm{Sym},#1}}

\newcommand{\din}[0]{d_{\mathrm{in}}}
\newcommand{\dinMean}[0]{\bar{d}_{\mathrm{in}}}
\newcommand{\dinM}[1]{d_{\mathrm{in},#1}}
\newcommand{\dmem}[0]{d_{\mathrm{mem}}}
\newcommand{\Vout}[0]{V_{\mathrm{out}}}
\newcommand{\VoutM}[1]{V_{\mathrm{out},#1}}
\newcommand{\Vin}[0]{V_{\mathrm{in}}}
\newcommand{\VinM}[1]{V_{\mathrm{in},#1}}
\newcommand{\VIN}[0]{\mathcal{V}_{\mathrm{in}}}
\newcommand{\VOUT}[0]{\mathcal{V}_{\mathrm{out}}}
\newcommand{\VOUTtot}[0]{\mathcal{V}_{\mathrm{out,tot}}}
\newcommand{\Vouttot}[0]{V_{\mathrm{out,tot}}}
\newcommand{\VINM}[1]{\mathcal{V}_{\mathrm{in},#1}}
\newcommand{\VOUTM}[1]{\mathcal{V}_{\mathrm{out},#1}}

\newcommand{\step}[0]{\mathrm{\Delta}}

\newcommand{\attCONST}[0]{\vartheta_{\mathrm{buf}}}
\newcommand{\attSec}[1]{\vartheta_{\mathrm{buf}}(#1)}

\newcommand{\attMean}{\bar{\vartheta}_{\mathrm{buf}}}

\newcommand{\ChinSec}[1]{C^{\mathrm{H^{+}}}_{\mathrm{in}}(#1)}
\newcommand{\CsinSec}[1]{C^{\Sub}_{\mathrm{in}}(#1)}
\newcommand{\ChinSwitch}[0]{C^{\mathrm{H^{+}}}_{\thresh}}
\newcommand{\ChoutSwitch}[0]{C^{\mathrm{H^{+}}}_{\mathrm{out},\thresh}}
\newcommand{\CiinSwitch}[1]{C^{\I}_{\thresh,m}}
\newcommand{\ChinEq}[0]{C^{\mathrm{H^{+}}}_{\mathrm{in,eq}}}
\newcommand{\ChinStart}[1]{C^{\mathrm{H^{+}}}_{\mathrm{in},0}(#1)}
\newcommand{\CsinStart}[1]{C^{\Sub}_{\mathrm{in},0}(#1)}

\newcommand{\tji}[2]{t^{(#1)}_{#2}}
\newcommand{\tj}[1]{t^{(#1)}}
\newcommand{\tsp}[1]{\tji{1}{#1}}
\newcommand{\tspNoi}[0]{\tj{1}}

\newcommand{\tssym}[1]{\tji{2}{#1}}
\newcommand{\tssymNoi}[0]{\tj{2}}
\newcommand{\tssymM}[2]{t^{\mathrm{s}}_{\mathrm{Sym},#1,#2}}
\newcommand{\tssymTheo}[1]{\tau^{\mathrm{s}}_{#1}}

\newcommand{\tep}[1]{\tji{3}{#1}}
\newcommand{\tepNoi}[0]{\tj{3}}
\newcommand{\tesym}[1]{\tji{4}{#1}}
\newcommand{\tesymNoi}[0]{\tj{4}}
\newcommand{\tsymTheoX}[2]{\tau^{#1}_{#2}}
\newcommand{\tsymX}[2]{t^{(#1)}_{#2}}
\newcommand{\tpX}[2]{t^{#1}_{\mathrm{P,#2}}}
\newcommand{\tSecStart}[1]{\tau(#1)}
\newcommand{\tActiveSym}[1]{t_{\mathrm{Sym}}^{\mathrm{a}}(#1)}
\newcommand{\tdep}[0]{t^{\mathrm{d}}_{\mathrm{Sym}}}

\newcommand{\lamW}[1]{\mathrm{W}\left \{ #1 \right \}}

\newcommand{\Chplus}[0]{C^{\Hplus}}
\newcommand{\ChplusX}[2]{C^{\Hplus}_{#1}(#2)}
\newcommand{\ka}[0]{k_{\mathrm{D}}}
\newcommand{\kPlus}[0]{k_{\mathrm{+}}}
\newcommand{\kMinus}[0]{k_{\mathrm{-}}}
\newcommand{\pKa}[0]{\mathrm{p}K_{\mathrm{D}}}

\newcommand{\Ci}[0]{C^{\I}}
\newcommand{\Cl}[0]{C^{\L}}
\newcommand{\Cil}[0]{C^{\I \L}}

\newcommand{\DinB}[0]{\Delta_{\mathrm{in}}}
\newcommand{\dinB}[0]{\delta_{\mathrm{in}}}
\newcommand{\DoutB}[0]{\Delta_{\mathrm{out}}}
\newcommand{\doutB}[0]{\delta_{\mathrm{out}}}
\newcommand{\DX}[1]{\Delta_{#1}}
\newcommand{\dX}[1]{\delta_{#1}}

\newcommand{\der}[0]{\mathrm{d}}

\newcommand{\iie}[1]{i^{\mathrm{I}}_{\mathrm{E}}(#1)}
\newcommand{\iieM}[2]{i^{\mathrm{I}}_{\mathrm{E},#1}(#2)}
\newcommand{\iir}[1]{i^{\mathrm{I}}_{\mathrm{R}}(#1)}
\newcommand{\iirM}[2]{i^{\mathrm{I}}_{\mathrm{R},#1}(#2)}
\newcommand{\isr}[1]{i^{\Sub}_{\mathrm{R}}(#1)}
\newcommand{\isrM}[2]{i^{\Sub}_{\mathrm{R},#1}(#2)}
\newcommand{\iil}[1]{i^{\mathrm{I}}_{\mathrm{L}}(#1)}
\newcommand{\iilM}[2]{i^{\mathrm{I}}_{\mathrm{L},#1}(#2)}
\newcommand{\iieNot}[0]{i^{\mathrm{I}}_{\mathrm{E}}}
\newcommand{\iirNot}[0]{i^{\mathrm{I}}_{\mathrm{R}}}
\newcommand{\isrNot}[0]{i^{\Sub}_{\mathrm{R}}}
\newcommand{\iilNot}[0]{i^{\mathrm{I}}_{\mathrm{L}}}
\newcommand{\iieNotM}[1]{i^{\mathrm{I}}_{\mathrm{E},#1}}
\newcommand{\iirNotM}[1]{i^{\mathrm{I}}_{\mathrm{R},#1}}
\newcommand{\isrNotM}[1]{i^{\Sub}_{\mathrm{R},#1}}
\newcommand{\iilNotM}[1]{i^{\mathrm{I}}_{\mathrm{L},#1}}
\newcommand{\Ni}[0]{N^{\mathrm{I}}}
\newcommand{\NiM}[1]{N^{\mathrm{I}}_{#1}}

\newcommand{\nves}[0]{n_{\mathrm{ves}}}
\newcommand{\nEx}[0]{n_{\mathrm{exp}}}
\newcommand{\nMod}[0]{n_{\mathrm{mod}}}
\newcommand{\probPump}[0]{p_\mathrm{P}}

\renewcommand{\Pr}[1]{\mathrm{Pr}\{#1\}}

\newcommand{\rTX}[0]{r_\mathrm{TX}}
\newcommand{\rRX}[0]{r_\mathrm{RX}}
\newcommand{\Diff}[0]{D_\mathrm{SM}}
\newcommand{\gsl}[0]{\gamma^{\Sub}_{\mathrm{L}}}
\newcommand{\Tbit}[0]{T_\mathrm{b}}
\newcommand{\erfc}[1]{\mathrm{erfc}\{#1\}}
\renewcommand{\exp}[1]{\mathrm{exp}\{#1\}}
\newcommand{\Tload}[0]{T_\mathrm{l}}
\newcommand{\TloadMin}[0]{T_\mathrm{l}^\mathrm{min}}
\newcommand{\TloadApprox}[0]{T_\mathrm{l}^\mathrm{app}}
\newcommand{\TloadMax}[0]{T_\mathrm{l}^\mathrm{max}}
\newcommand{\Tdeload}[0]{T_\mathrm{d}}
\newcommand{\TdeloadMin}[0]{T_\mathrm{d}^\mathrm{min}}
\newcommand{\TdeloadApprox}[0]{T_\mathrm{d}^\mathrm{app}}
\newcommand{\TdeloadMax}[0]{T_\mathrm{d}^\mathrm{max}}
\newcommand{\todo}[1]{\textcolor{red}{#1}}
\newcommand{\KmH}[0]{\mathrm{K}_\mathrm{m,H^+}}
\newcommand{\Hill}[0]{n}
\newcommand{\ChinThresh}[0]{C^{\Hplus}_{\mathrm{in}, \xi}}
\newcommand{\ChinS}[0]{C\mathrm{^{\Hplus}_{in,s}}}
\newcommand{\ChinE}[0]{C\mathrm{^{\Hplus}_{in,e}}}
\newcommand{\NSM}[0]{N_\mathrm{SM}}
\newcommand{\rSM}[0]{\bar{n}_\mathrm{SM}}
\newcommand{\rN}[0]{\bar{n}_\mathrm{N}}
\newcommand{\TGuard}[0]{T_\mathrm{g}}
\newcommand{\isl}[1]{i^{\Sub}_{\mathrm{L}}(#1)}
\newcommand{\VRX}[0]{V_\mathrm{RX}}
\newcommand{\Rate}[0]{R_\mathrm{T}}
\newcommand{\lOOK}[1]{l_{\mathrm{OOK}}(#1)}
\newcommand{\pOOK}[1]{p_{\mathrm{OOK}}(#1)}
\newcommand{\lOOKGI}[1]{l_{\mathrm{OOK-GI}}(#1)}
\newcommand{\pOOKGI}[1]{p_{\mathrm{OOK-GI}}(#1)}
\newcommand{\lMAM}[1]{l_{\mathrm{MAM}}(#1)}
\newcommand{\pMAM}[1]{p_{\mathrm{MAM}}(#1)}
\newcommand{\lMEM}[1]{l_{\mathrm{MEM}}(#1)}
\newcommand{\pMEM}[1]{p_{\mathrm{MEM}}(#1)}
\renewcommand{\vec}[1]{\mathbf{#1}}
\newcommand{\Nsout}[1]{N^{\mathrm{SM}}(#1)}

\begin{abstract}
\Ac{MC} enables information exchange through the transmission of \acp{SM} and holds promise for many innovative applications. However, most existing works in \ac{MC} rely on simplified \ac{TX} models that do not account for the physical and biochemical limitations of realistic biological hardware and environments. This work extends previous efforts toward developing models for practical \ac{MC} systems by proposing a more realistic TX model that incorporates the delay in \ac{SM} release and \ac{TX} noise introduced by biological components. Building on this more realistic, functionalized vesicle-based \ac{TX} model, we propose two novel modulation schemes specifically designed for this \ac{TX} to mitigate \ac{TX}-induced memory effects that arise from delayed and imperfectly controllable \ac{SM} release. The proposed modulation schemes enable low-complexity \acl{RX} designs by mitigating memory effects directly at the \ac{TX}. Numerical evaluations demonstrate that the proposed schemes improve communication reliability under realistic biochemical constraints, offering an important step toward physically realizable \ac{MC} systems.
\end{abstract}

\acresetall

\section{Introduction}
\noindent
\Ac{MC} is a bio-inspired communication paradigm in which \acp{SM} serve as information carriers \cite{Nakano2013}. \ac{MC} shall enable communication in environments where conventional communication approaches are impractical and envisions applications in health monitoring, targeted drug delivery, and agriculture \cite{Felicetti2016,Chude-Okonkwo2017,Tuccitto2024}. Despite its promise, most existing \ac{MC} research remains theoretical and is often detached from practical realization, making oversimplifying assumptions that neglect physical and biochemical limitations of biological hardware. Consequently, there is a lack of accurate models for practically realizable \acp{TX} and \acp{RX}, which are essential for the practical implementation of \ac{MC} systems. 
\noindent Realistic \ac{TX} models must account for the effects imposed by biological hardware, including delay, buffering, and imperfect control of \ac{SM} release. \textit{Delay} is caused by the finite reaction rates of the cascades of biochemical processes that are facilitated by biological components \cite{Bamann2014}. \textit{Buffering effects}, which are expected in many \ac{MC}-relevant biological environments such as the bloodstream \cite{Ellison1958}, further impact these reaction dynamics. Moreover, \acp{SM} may leak from the \ac{TX} or be released in an uncontrolled manner, resulting in undesired \textit{\ac{TX} noise}.
Several efforts have been made toward more realistic \ac{TX} designs. In \cite{Schaefer2022}, a vesicle-based \ac{TX} was proposed, where \ac{SM} release is controlled by pH-induced changes in membrane permeability. In \cite{Huang2022}, a membrane-fusion-based \ac{TX} inspired by natural exocytosis processes was proposed. Recently, in \cite{Jing2024}, an imperfect molecule shift keying \ac{TX} that accounts for energy consumption during \ac{SM} replenishment was investigated. In our previous work \cite{tomDieck2025}, we presented a realistic, functionalized vesicle-based \ac{TX} model incorporating ion pumps as biological modules for energy generation via ion gradients, and symporters as biological modules for \ac{SM} release. The energy module can be externally controlled by light, while the ion gradient generated by this module governs the activity of the symporters. As these processes are cascaded, i.e., this \ac{TX} must first establish a concentration gradient before \acp{SM} are released, it inherently causes a delay between the light signal and the \ac{SM} release, which leads to memory effects at the \ac{TX}. Additionally, in \cite{tomDieck2025}, the impact of monoprotic pH buffers was considered, which further influence the transport processes of ions and, thus, induce additional memory effects. In this work, we extend the \ac{TX} model proposed in \cite{tomDieck2025} by introducing a more realistic model for \ac{SM} release. Specifically, the release of \acp{SM} in the envisioned vesicle-based \ac{TX} cannot be perfectly controlled, as unintended \ac{SM} leakage through the vesicle membrane and undesired release by the symporters may occur.
Because of their specific properties, realistic \ac{TX} models require tailor-made modulation and detection schemes. For example, delays introduced at the \ac{TX} by biological hardware induce memory deteriorating communication performance. Typically, memory in \ac{MC} systems is attributed to \ac{ISI} caused by stochastic \ac{SM}  propagation. Common mitigation strategies include equalization \cite{Tepekule2015}, sequence estimation \cite{Kilinc2013}, and channel coding \cite{Lu2015}. However, such methods often reduce throughput and/or demand significant computational resources at the \ac{RX} side.
\begin{figure}
    \centering
    \includegraphics[width=0.48\textwidth]{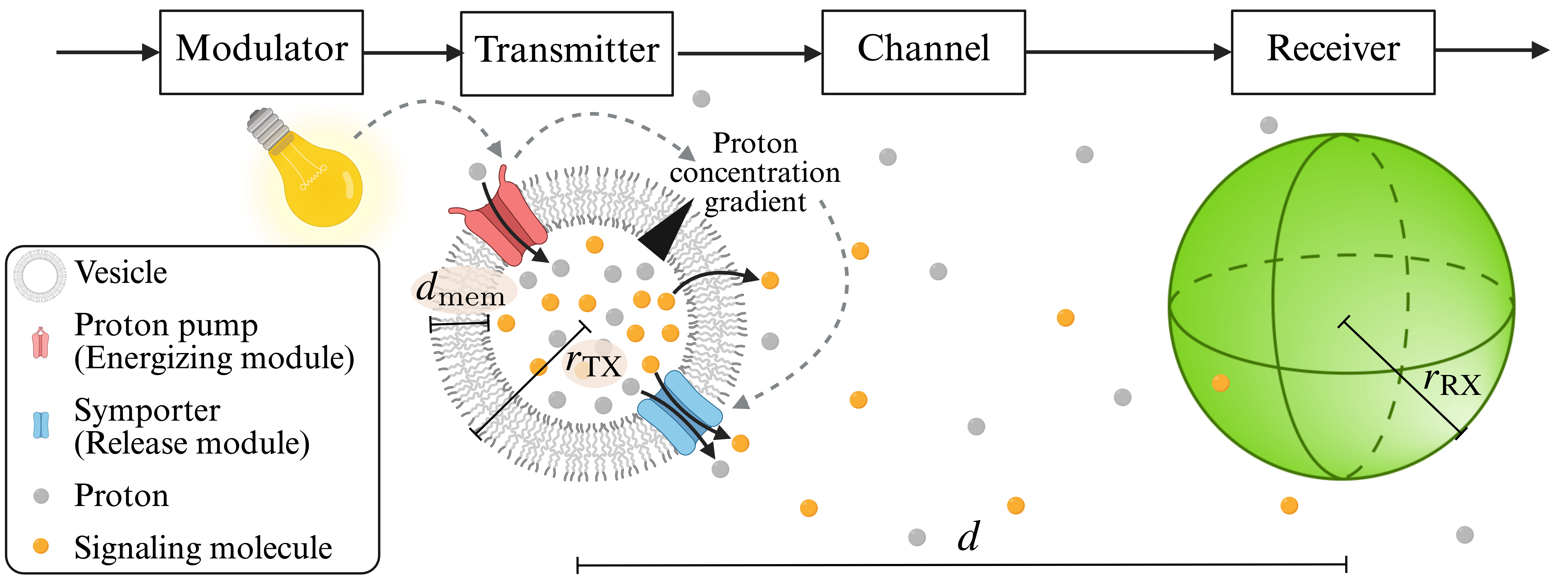}
    \vspace{-0.5em}
    \caption{Schematic illustration of the considered end-to-end MC system featuring a binary \acs{LED} as the modulator, the realizable \ac{TX}, a diffusive channel, and a \acf{FAR} (from left to right, not drawn to scale). Created with BioRender.com.}
    \vspace{-1em}
    \label{fig:comm_system}
\end{figure}
\noindent In contrast, in some \ac{MC} systems, memory effects can be mitigated directly at the \ac{TX} side, thereby avoiding any increase in \ac{RX} complexity. A practical strategy is to employ an \ac{ISI}-aware modulation scheme, which can often be easily adapted by the user in \ac{MC} systems \cite{tomDieck2025,scherer2025closed,Wietfeld2024}. Several modulation schemes for \ac{ISI} mitigation have been proposed in the literature (see \cite{kuran2020survey} for an overview); however, most require modifications at the \ac{RX}. In contrast, this work presents two novel modulation schemes tailor-made for realistic biological \ac{TX} hardware. These schemes tackle the memory effects inherent to realizable \acp{TX} on the \ac{TX} side such that an unaltered low-complexity \ac{RX} can be used. Such an approach is especially useful when the \ac{RX} is set or cannot be tuned adequately, as is assumed here.
The main contributions of this paper are as follows:

\begin{itemize}
\item We extend the realizable \ac{TX} model from \cite{tomDieck2025} to account for the kinetics of the release module more accurately and to include \ac{TX} noise caused by \acp{SM} leakage and the soft activation of the release module.
\item We propose two novel modulation schemes for mitigating \ac{TX}-induced memory effects.
\item We compare the proposed modulation schemes to conventional benchmark schemes and analyze their performance in different system settings.
\end{itemize}

\noindent The remainder of this paper is organized as follows. \Cref{sec:system} introduces the system model, including the model for the release kinetics. The novel modulation schemes are proposed in \cref{sec:modulation}. Simulation results and discussions are presented in \cref{sec:results}, and conclusions with directions for future work are provided in \cref{sec:conclusion}.

\section{Communication System}
\label{sec:system}

\begin{figure*}
  \centering
  \begin{minipage}[t]{0.49\textwidth}
    \centering
    \includegraphics[width=\linewidth]{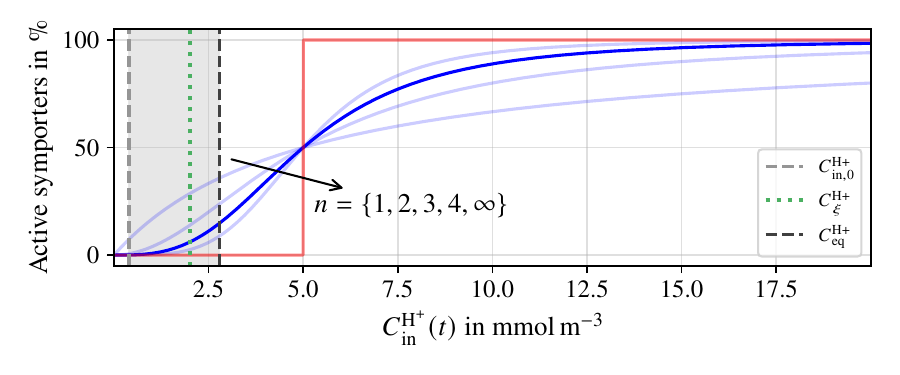}
    \vspace{-2em}
    \caption{Hill curve for $n \in \{1, 2, 3, 4, \infty\}$. The curve for $n \rightarrow \infty$ (hard threshold) is highlighted in red and the activity threshold $\ChinSwitch$ for $n = 3$ (see \cref{sec:mam}) is indicated by a green dotted line. $\ChinZERO$ and $\ChinEq$ are indicated by gray dashed lines and the gray-shaded area indicates the operation interval of the \ac{TX}. The parameter values in \cref{tab:params} were used.}
    \label{fig:hill}
    \vspace{-1em}
  \end{minipage} \hfill
  \begin{minipage}[t]{0.49\textwidth}
  \centering
  \includegraphics[width=\linewidth]{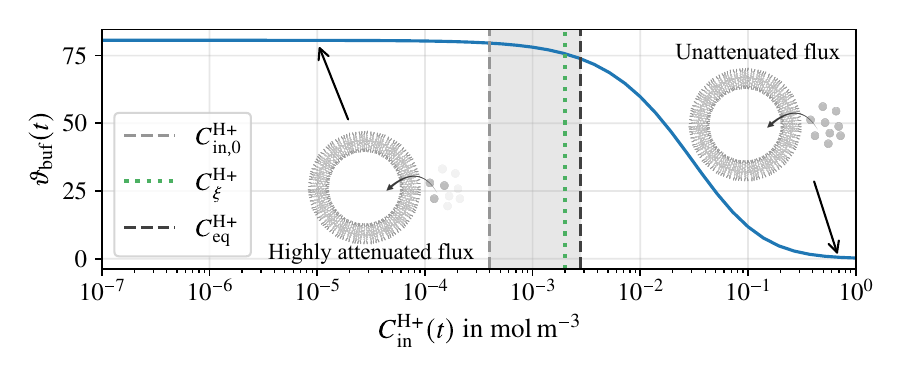}
    \vspace{-2em}
    \caption{Flux attenuation factor $\attSec{t}$ vs. intravesicular $\Hplus$ concentration for the system parameters in \cref{tab:params} and buffer molarity $\CZERO = \SI{5}{\mol \per \cubic \meter}$. The green dotted line corresponds to $\ChinThresh$. $\ChinZERO$ and $\ChinEq$ are indicated by gray dashed lines. The gray-shaded area indicates the operation interval of the \ac{TX}. Created in part with BioRender.com.}
    \label{fig:buffer}
    \vspace{-1em}
  \end{minipage}
\end{figure*}

\noindent The considered end-to-end \ac{MC} system is illustrated in \cref{fig:comm_system}. It consists of a modulator, which creates a physical signal (i.e., light), and the \ac{TX}, which is a spherical nanoscale biological device (i.e., a vesicle) with radius $\rTX$ and membrane thickness $\dmem$. All components besides the modulator are immersed in aqueous buffer solution. The communication channel is modeled as a free space diffusive environment in which \acp{SM} propagate with diffusion coefficient $\Diff$, as the extravesicular volume is much larger than the \ac{TX} volume \cite{tomDieck2025}. A \acf{FAR} of radius $\rRX$ is considered, separated from the \ac{TX} by a center-to-center distance $d$.

\subsection{Modulator}
\noindent The modulator is an external \acf{LED} (see \cref{fig:comm_system}), which has binary states: ON and OFF. Hence, also the \ac{LED} excitation signal $l(t)$, where $t$ denotes the time, is binary, i.e., $l(t) \in \{0, 1\}$. We assume that the switching times between the \ac{LED} states can be chosen arbitrarily according to the modulation scheme.

\subsection{Realistic Transmitter Model and Properties}
\noindent As \ac{TX} we consider the spherical, vesicle-based nanodevice previously proposed in  \cite{tomDieck2025}.

\subsubsection{Functionality of the Transmitter}
The vesicle-based \ac{TX} contains \acp{SM} as cargo inside. It facilitates \ac{SM} release via a cascade of biochemical reactions. First, a light-driven proton-pump (energizing module, depicted in red in \cref{fig:comm_system}\footnote{Generally, the energizing and release module consist of multiple proteins each. For sake of simplicity, this is not shown in \cref{fig:comm_system}.}), which is controlled by the modulator, is used to transport $\Hplus$ ions (gray circles in \cref{fig:comm_system}) into the vesicle, resulting in a change of the intravesicular $\Hplus$ concentration, $\Chin{t}$.
The resulting $\Hplus$ flux over the membrane, $\ihp{t}$, is given by \cite[Eq. (6)]{tomDieck2025}
\begin{equation}
    \ihp{t} = \Chout{t}(\ChoutZERO )^{-1} \ghp l(t),
    \label{eq:flux_p}
\end{equation}
where $\Chout{t}$, $\ChoutZERO$, and $\ghp$ denote the extravesicular $\Hplus$ concentration at time $t$, the extravesicular $\Hplus$ concentration at $t = 0$, and the effective pumping rate constant of the energizing module in \si{\mol \per \second}, respectively. 
The resulting concentration gradient of $\Hplus$ across the membrane serves as energy supply for a second active transport protein, the $\Hplus$/\ac{SM} symporter (release module, depicted in blue in \cref{fig:comm_system}). In particular, the symporter exploits the transport of $\Hplus$ ions along the concentration gradient to simultaneously transport \acp{SM}. The flux of $\Hplus$ and \acp{SM} across the vesicle membrane caused by the symporters will be detailed in \cref{sec:details_soft_activation}.
The considered vesicle-based \ac{TX} model implicitly captures the delay of \ac{SM} release that occurs before the concentration gradient of the protons becomes large enough to power the release module. The delayed response of the \ac{TX} to the external stimulus is illustrated by the cascade of gray dashed arrows in \cref{fig:comm_system}. Delay effects are to be expected in many realizable \acp{TX} for \ac{MC} systems, as mentioned in, e.g., \cite{Schaefer2022,Huang2022}, and therefore must be accounted for in \ac{MC} system design. In addition to cascaded reactions (i.e., conformational state changes of the transport proteins), the vesicle-based \ac{TX} model accounts for leakage of both $\Hplus$ and \acp{SM}, where the respective molecules diffuse through the vesicle membrane passively along their concentration gradients.
\\ \noindent Due to the nature of the fluxes across the membrane (discussed in detail below), the \ac{TX} operates in a bounded range of intravesicular $\Hplus$ concentrations. For the design of the modulation scheme, this concentration interval has to be taken into consideration. In case of the proposed \ac{TX}, this interval ranges between the minimum $\Hplus$ concentration, $\ChinZERO$, and the maximum $\Hplus$ concentration, $\ChinEq$ (see dashed lines and gray-shaded areas in \cref{fig:hill,fig:buffer}). $\ChinZERO$ and $\ChinEq$ are the concentrations at which the net flux of $\Hplus$ over the vesicle membrane, for which the formulas are given below, is zero, given the energizing module is inactive ($\ChinZERO$) and active ($\ChinEq$), respectively. The value of $\ChinEq$ can be obtained analytically under the assumption that the \ac{SM} concentration is constant.
The derivation of $\ChinEq$ for $n = 3$ can be found in \cref{app:B}.

\subsubsection{Soft Activation of the Symporters}
\label{sec:details_soft_activation}

While the model in \cite{tomDieck2025} assumed a hard threshold for symporter activation, we extend \cite{tomDieck2025} and adopt a more realistic model by assuming a soft activation of the symporters. In particular, we describe the release kinetics of the symporters using Hill kinetics (shown in  \cref{fig:hill}) as proposed in \cite{Lolkema2015}. With this extension, the flux of $\Hplus$ caused by the symporters, $\ihsym{t}$, is given as follows

\begin{equation}
    \ihsym{t} = \ghsym \; \frac{\Csin{t}}{\Csin{t} + \Km} \;\frac{(\Chin{t})^{n}}{(\Chin{t})^{n} + \KmH^{n}},
    \label{eq:hill}
\end{equation}

\noindent with the last fraction in \eqref{eq:hill} corresponding to the Hill kinetics and $\ghsym$, $\Csin{t}$, $\Km$, $\KmH$, and $n$ denoting the $\Hplus$ transport rate constant of the symporters, the intravesicular $\Sub$ concentration, the Michaelis-Menten constant, the $\Hplus$ concentration resulting in $50 \, \%$ symporter activation (in case that all symporters have bound \acp{SM}), and the Hill coefficient, respectively. The outward flux of \acp{SM} caused by the symporters is $\issym{t} = \ihsym{t} / \fSym$, where $\fSym$ denotes the ratio of the number of $\Hplus$ molecules transported per transported \ac{SM}. For $n \rightarrow \infty$, the Hill kinetics correspond to a hard threshold, as used in \cite{tomDieck2025}. For finite $n$, however, the symporters constantly release \acp{SM}, with a transport rate depending on $\Chin{t}$. In addition to passive diffusion of \acp{SM} through the membrane, the slow release of \acp{SM} due to the soft activation of the symporters can be regarded as a form of \ac{TX} noise.

\subsubsection{Buffering Effects}

The \ac{TX} model accounts for buffering effects. In the biological context, buffers are solutions capable of resisting changes in $\Hplus$ concentration. The presence of such buffers ensures that the pH value remains stable even when pH-altering reactions occur. Buffers are typically present in biological environments, and thus their effect must be considered in \ac{MC} systems employing $\Hplus$ (or other ions). In such systems, buffers can influence system dynamics considerably, as investigated in \cite{tomDieck2025}. In \cite[Eq. (21)]{tomDieck2025}, the influence of the buffer is captured by a flux attenuation factor $\attSec{t} \approx \ka \CZERO(\Chin{t} + \ka)^{-2}$ for the $\Hplus$ flux over the vesicle membrane (as illustrated by the embedded drawings in \cref{fig:buffer}), where $\ka$ and $\CZERO$ denote the dissociation constant of the buffer in \si{\mol \per \cubic \meter} and the buffer molarity in \si{\mol \per \cubic \meter}, respectively. The presence of the buffer can have a significant impact on the operation of the \ac{TX}, depending on the range of attenuation factors that occur in a given system. In \cref{fig:buffer}, the changes in $\attSec{t}$ for varying $\Chin{t}$ for the parameters in \cref{tab:params} are visualized. We observe that in the relevant interval between $\ChinZERO$ and $\ChinEq$ the attenuation factor $\attSec{t}$ is large, which suggests that buffering effects can not be neglected in models for the considered system.

\subsubsection{Leakage}

Synthetic vesicles, also those with polymeric membranes, are never perfectly impenetrable and, thus, lose some of their cargo due to leakage of molecules over their membranes \cite{Rideau2018}. We consider the membrane to be permeable to $\Hplus$ with permeability coefficient $\ghl$, resulting in leakage flux $\ihl{t} = \ghl \left ( \Chin{t} - \Chout{t} \right )$ (see \cite[Eq. (8)]{tomDieck2025}). Additionally, we consider the membrane of the \ac{TX} to be permeable to \acp{SM} with permeability coefficient $\gsl$. As proposed in \cite{Noel2014}, this effect of undesired \acp{SM} release can be regarded as \ac{TX} noise in \ac{MC} systems. Analogous to the $\Hplus$ leakage flux, we define the \ac{SM} leakage flux as $\isl{t} = \gsl \left(\Csin{t} - \Csout{t}\right)$, where $\Csout{t}$ denotes the extravesicular $\Sub$ concentration. The \ac{SM} leakage flux varies with time as the \ac{SM} supply in the vesicle diminishes.

\noindent To summarize, there are several effects that occur in practical \acp{TX}, which the proposed model captures. Due to their finite rates, the \textbf{cascaded reactions} facilitating the release of \acp{SM} implicitly add delay to the release signal. The \textbf{buffer} directly influences changes in $\Chin{t}$, thus, it also influences the delay of \acp{SM} release. Additionally, the \textbf{\ac{SM} leakage} and \textbf{soft activation of the symporters} lead to \ac{TX} noise.

\subsection{Modeling and Simulation Methods}

\noindent In \cite{tomDieck2025}, approximate analytical models for the dynamics of $\Chin{t}$ and $\Csin{t}$ were proposed. Additionally, an exact numerical \ac{FDM} was used as a baseline. As we further extended the model from \cite{tomDieck2025} to a soft activation of the symporters, we use \ac{FDM} to adequately model the \ac{TX} behavior in this work. We then convolve the number of released \acp{SM} over time obtained numerically from the \ac{TX} model, $\Nsout{t}$, with the impulse response of the channel, $h(t)$ (see \cite[Eq. (9)]{Huang2022}), to obtain the expected received signal, i.e., $r(t) = h(t) \ast \Nsout{t}$, where $\ast$ denotes the convolution operator.

\vspace{-0.15cm}
\section{Modulation Schemes}
\label{sec:modulation}
\begin{figure}
    \centering
    \includegraphics[width=0.48\textwidth]{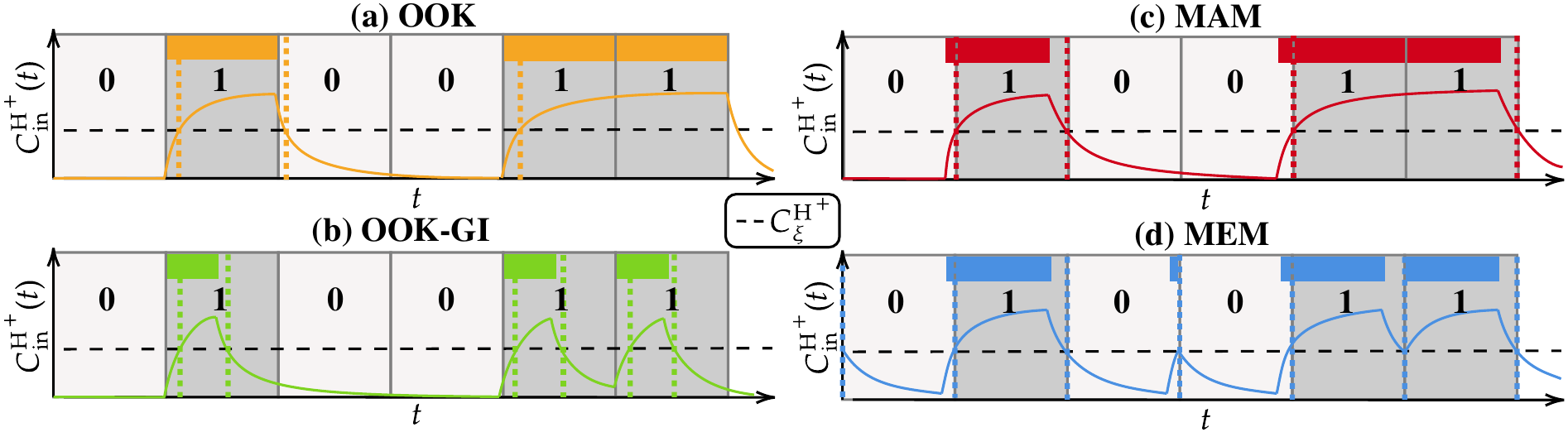}
    \vspace{-1em}
    \caption{$\Chin{t}$ for the considered modulation schemes and an exemplary bit sequence. Gray-shaded areas indicate bit ``1'' intervals. (a) \ac{OOK}, (b) \ac{OOK-GI}, (c) \ac{MAM}, (d) \ac{MEM}. Colored bars indicate times during which the \ac{LED} is switched on for the different modulation schemes and the horizontal black dashed line indicates $\ChinThresh$. Vertical dotted colored lines indicate the times at which $\ChinThresh$ is crossed.}
    \label{fig:modulation}
    \vspace{-2em}
\end{figure}

\noindent In this section, we introduce the different modulation schemes examined in the proposed MC system. After briefly reviewing two well-investigated benchmark modulation schemes, namely \ac{OOK} and \ac{OOK-GI}, we propose two novel memory-mitigating modulation schemes, namely the \ac{MAM} scheme and \ac{MEM} scheme. We derive \ac{LED} signals for all modulation schemes for a vector of binary symbols $\vec{b} \in \{0,1\}^{L}$, where $L \in \mathbb{N}$ and $\mathbb{N}$ denote the length of the sequence of binary symbols and the set of natural numbers, respectively. The $i$-th element of $\vec{b}$ is denoted $b_i$, where $i \in \{0, \dots, L - 1\}$. All four considered modulation schemes are illustrated in \cref{fig:modulation}. 

\subsection{On-Off Keying}
\noindent \ac{OOK} is a common modulation scheme, widely used in various communication systems, as it is straightforward and universally applicable. In our system, \ac{OOK} is realized by continuous light illumination during the entire bit ``1'' intervals of length $\Tbit$, and no illumination during bit ``0'' intervals (see \cref{fig:modulation}a), i.e., the \ac{LED} signal for \ac{OOK}, $\lOOK{t}$, becomes
\begin{equation}
    \lOOK{t} = \sum_{i = 0}^{L - 1} b_i \pOOK{t - i\Tbit}, \quad \pOOK{t} = \mathds{1}_{[0, \Tbit)}(t),
\end{equation}

\noindent where $\pOOK{t}$ and $\mathds{1}_\mathcal{X}(x)$ denote respectively the pulse shape for a single bit and the indicator function, which is 1 if $x \in \mathcal{X}$ (i.e., if $x$ is an element of set $\mathcal{X}$) and 0 otherwise.
In the context of the proposed \ac{TX}, \ac{OOK} is subject to the delay in the release of \acp{SM} caused by cascaded biochemical reactions. This can be seen in \cref{fig:modulation}a by the misalignment of the dotted orange lines (indicating when the symporters become (in)active, see \cref{sec:thresholding}) and the beginning and end of the bit intervals.
\subsection{On-Off Keying with Guard Interval}
\noindent One possible approach for reducing memory effects and \ac{ISI} at the \ac{TX} is the use of a \ac{GI}, which we denote by \ac{OOK-GI}. The \acs{GI} ensures a fixed period of time $\TGuard < \Tbit$ of non-excitation at the end of each bit interval (see \cref{fig:modulation}b). Therefore, the \ac{LED} signal for \ac{OOK-GI}, $\lOOKGI{t}$, reads
\begin{equation}
 \lOOKGI{t} = \sum_{i = 0}^{L - 1} b_i \pOOKGI{t - i\Tbit},
\end{equation}
\noindent where $\pOOKGI{t} = \mathds{1}_{[0, \Tbit  - \TGuard)}(t)$ denotes the pulse shape for a single bit for \ac{OOK-GI}.
The \ac{GI} gives the memory-prone system time to re-establish the initial equilibrium state $\ChinZERO$, such that the \ac{TX} ideally produces the same release signal independent of the previously transmitted sequence. Therefore, the \ac{GI} is expected to improve the communication reliability in comparison to \ac{OOK}. To achieve this effect, it is crucial to adequately choose $\TGuard$, as a large \ac{GI} comes at the cost of reduced received signal strength. Here, we assume $\TGuard = 0.5 \Tbit$.
\subsection{Memory-Aware Modulation}
\label{sec:mam}

\noindent As the realizable \ac{TX} exhibits a delay in \ac{SM} release, a modulation scheme could be adapted such that it compensates for this delay. The goal is to shape the modulation signal such that $\Chin{t}$ reaches a predefined threshold, $\ChinThresh$, at the start of the next bit interval whenever a bit switch occurs, such that the \ac{TX} is in an ``active'' state at the beginning of bit ``1''s and in an ``inactive'' state at the beginning of bit ``0''s. In \cref{fig:modulation}, this is illustrated by the alignment between the shifted red bar (\ac{LED}/modulation signal) and the bit interval boundaries. With this adaptation, the system can immediately produce the desired response (i.e., strong \acp{SM} release during bit “1” intervals) at the start of a new bit ``1'' interval. We refer to this approach as \acf{MAM}.

\subsubsection{Thresholding}
\label{sec:thresholding}
As we adopt a realistic soft thresholding model for the symporters, the question arises how to define their ``active'' and ``inactive'' states. We observe in \cref{fig:hill}, that for a specific Hill coefficient $\Hill$, there are $\Chin{t}$ values at which the rate changes more quickly upon an increase in $\Hplus$. We define this concentration, $\ChinThresh$, as a threshold. In particular, we define $\ChinThresh$ to be the concentration at which the \emph{slope} of the Hill curve changes most rapidly. To obtain $\ChinThresh$, we take the third derivative of \eqref{eq:hill} and set it to zero, leading to

\begin{equation}
    \ChinThresh = \left(\frac{-2 \KmH^\Hill (\Hill^2 - 1) \pm \KmH^\Hill \Hill \sqrt{3(\Hill^2 - 1)}}{2 + 3 \Hill + \Hill^2}\right)^{\frac{1}{\Hill}},
\end{equation}

\noindent where the resulting $\ChinThresh$ must be nonnegative and real-valued. $\ChinThresh$ is shown in \cref{fig:hill} for $\Hill = 3$ as a green dotted line.

\begin{figure}[t]
    \centering
    \includegraphics[width=0.48\textwidth]{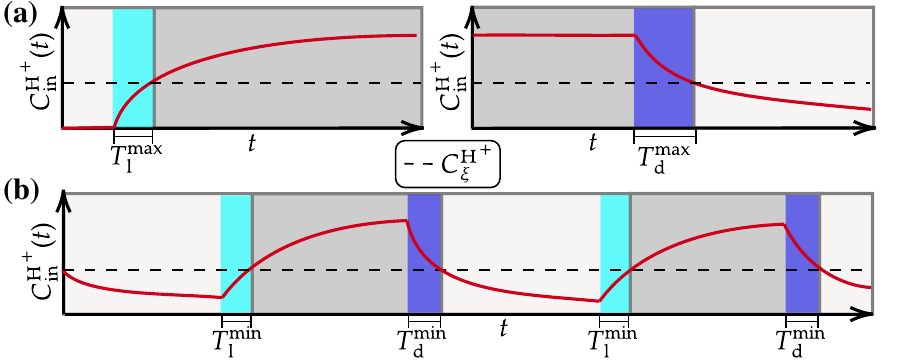}
    \vspace{-1em}
    \caption{(a) Derivation of upper bounds for $\Tload$ and $\Tdeload$, i.e., $\TloadMax$ and $\TdeloadMax$. (b) Derivation of lower bounds for $\Tload$ and $\Tdeload$, i.e., $\TloadMin$ and $\TdeloadMin$.}
    \vspace{-2em}
    \label{fig:bounds}
\end{figure}

\subsubsection{Loading and Deloading Times}
\label{sec:loadinganddeloading}

Implementing \ac{MAM} requires determining the loading duration $\Tload$ and deloading duration $\Tdeload$. To reduce the required computational effort, we opt to compute $\Tload$ and $\Tdeload$ \emph{offline} before transmission and do not vary them over time. $\Tload$ is the time the \ac{LED} must be in the ON state at the end of a bit ``0'' interval such that $\Chin{t}$ reaches $\ChinThresh$ by the start of the following bit ``1'' interval (shown in cyan in \cref{fig:bounds}a and b). $\Tdeload$ is the duration the \ac{LED} needs to be switched off at the end of a bit ``1'' interval such that $\Chin{t}$ decreases to $\ChinThresh$ at the start of the following bit ``0'' interval (shown in blue in \cref{fig:bounds}a and b). Since the calculations of $\Tload$ and $\Tdeload$ are analogous, we focus on $\Tload$ in the following. 
The loading time $\Tload$ is obtained by setting $\Chin{\Tload} = \ChinThresh$ and solving for $\Tload$. There are three caveats to consider when trying to obtain a value for $\Tload$: (C1) No closed-form solution for $\Chin{t}$ exists, making the derivation of an analytical expression for $\Tload$ infeasible. (C2) The $\Hplus$ concentration at the beginning of the $i$-th bit interval, $\Chin{i \Tbit}$, is unknown and depends on previous bit transmissions. (C3) The concentration of $\Hplus$ at the beginning of the loading interval, $\ChinS = \Chin{i\Tbit - \Tload}$ of the $i$-th bit where $b_i = 1$, is required for the calculation of $\Tload$ but depends on $\Tload$ itself. Due to (C1)--(C3), we opt to obtain numerical upper and lower bounds for $\Tload$, respectively (mitigating (C1) and (C2)). Additionally, for the lower bound, we use the block Gauss-Seidel method \cite{Saad2003} for obtaining $\ChinS$ from $\Tload$ and vice versa iteratively, until the values converge (mitigating (C3)).
Details for the proposed method can be found in \cref{app:A}.

\noindent The upper and lower bounds for $\Tload$ and $\Tdeload$ for \ac{MAM} can be derived as follows:

\begin{itemize}[leftmargin=0.3cm]
\item $\TloadMax$ and $\TdeloadMax$ denote the time required for $\Chin{t}$ to reach $\ChinThresh$ when the system starts at the \textit{minimum} $\Chin{t}$, i.e., $\ChinZERO$ (for $\TloadMax$), and at the \textit{maximum} $\Chin{t}$, i.e., $\ChinEq$ (for $\TdeloadMax$), respectively. Here, $\ChinS = \ChinZERO$ and $\ChinS = \ChinEq$ denote the respective initial proton concentrations. Hence, $\TloadMax$ and $\TdeloadMax$ represent upper bounds, which can be obtained from an \ac{FDM}, and are shown in \cref{fig:bounds}a (cyan and blue, respectively).
\item $\TloadMin$ and $\TdeloadMin$ are obtained by assuming alternating bit sequences (``0101...''), where $\Chin{i \Tbit}$ is --- due to the alternation --- always $\ChinThresh$. This approach underestimates $\Tload$ and $\Tdeload$ when no bit switch occurs. Hence, $\TloadMin$ and $\TdeloadMin$ constitute lower bounds, and are shown in \cref{fig:bounds}b (cyan and blue, respectively).
\end{itemize}
\noindent Finally, the \ac{LED} signal for \ac{MAM}, $\lMAM{t}$, is obtained as
\vspace{-0.1cm}
\begin{equation}
\begin{split}
    \lMAM{t} = & \sum_{i = 0}^{L - 2} {\begin{cases}
        b_i \pOOK{t - i \Tbit}, & \text{if } b_i = b_{i + 1}, \\
        \mathds{1}_{[0, \Tbit - \Tdeload)}(t - i \Tbit), & \text{if } b_i = 1, b_{i + 1} = 0, \\
        \mathds{1}_{[\Tbit - \Tload, \Tbit)}(t - i \Tbit), & \text{if } b_i = 0, b_{i + 1} = 1,   \end{cases}} \\
        & + b_{L-1} \pOOK{t - (L - 1)\Tbit}.
\end{split}
\end{equation}
\vspace{-0.1cm}

\subsection{Memory-Erasing Modulation}

\noindent The principle underlying \ac{MAM} can be extended to erase all memory induced by previous transmissions. We denote this modulation scheme as \ac{MEM} (see \cref{fig:modulation}d). The required extension ensures that $\Chin{t=i\Tbit} = \ChinThresh, \; \forall i \in \{0, ..., L -1\}$ is enforced at the beginning of all bit intervals --- in contrast to \ac{MAM}, which only takes effect for bit switches. The \ac{LED} signal for \ac{MEM}, $\lMEM{t}$, becomes
\vspace{-0.05cm}
\begin{equation}
    \lMEM{t} = \sum_{i = 0}^{L - 1} \begin{cases}
        \mathds{1}_{[0, \Tbit - \Tdeload)}(t - i \Tbit), & \text{if } b_i = 1, \\
        \mathds{1}_{[\Tbit - \Tload, \Tbit)}(t - i \Tbit), & \text{if } b_i = 0.
    \end{cases}
\end{equation}

This ensures that the \ac{TX} states at the beginning of all bit intervals are as similar as possible. In turn, \ac{MEM} exhibits more loading and deloading periods than \ac{MAM}, which increases the average number of \acp{SM} released during bit ``0''s and decreases the average number of \acp{SM} released during bit ``1''s. Hence --- depending on the detector --- \ac{MAM} can ease the distinguishability of the two symbols.

\section{Simulation Results}
\label{sec:results}

\begin{table}[!t]
    \vspace{0.5em}
    \centering
    \caption{Default parameters used.}
    \vspace{-1em}
    \def\arraystretch{1.4}
    \resizebox{\columnwidth}{!}{
    \begin{tabular}{|l|r@{\hspace{1pt}}l|c|}
        \hline
        Parameter & \multicolumn{2}{c|}{Value} & Ref.\\
        \hline
        \hline
        $\step t$ & $\num{1e-5}$ & $\si{\second}$ &  \\ \hline
        $\rTX$ & $\num{60}$ & $\si{\nano\meter}$ & \cite{tomDieck2025} \\ \hline
        $\rRX$ & $\num{200}$ & $\si{\nano\meter}$ & \cite{Rideau2018} \\ \hline
        $\Diff$ & $\num{9e-11}$ & $\si{\square\meter\per\second}$ & \\ \hline
        $d$ & $\num{2}$ & $\si{\micro\meter}$ &  \\ \hline
        $\CsinZERO$ & \num{300} & \si{\mol\per\cubic\meter} & \cite{tomDieck2025} \\ \hline
        $\ka$ & $\num{6.2e-5}$ & \si{\mol \per \cubic \meter} & \cite{tomDieck2025}  \\ \hline
        $\dmem$ & $\num{14}$ & $\si{\nano\meter}$ & \cite{tomDieck2025} \\ \hline
    \end{tabular}
    \begin{tabular}{|l|r@{\hspace{1pt}}l|c|}
        \hline
        Parameter & \multicolumn{2}{c|}{Value} & Ref.\\
        \hline
        \hline
        $\ChinZERO$, $\ChoutZERO$ & $\num{3.98e-4}$ & $\si{\mol\per\cubic\meter}$ &  \\ \hline
        $\Km$ & $\num{1.3e-2}$ & $\si{\mol\per\cubic\meter}$ & \cite{tomDieck2025} \\ \hline
        $\KmH$ & $\num{5e-3}$ & $\si{\mol\per\cubic\meter}$ &  \cite{Hummel2011} \\ \hline
        $n, \fSym$ & $\num{3}, \num{3}$ &  & \cite{Lolkema2015},\cite{tomDieck2025} \\ \hline
        $\gssym$ & $\num{1000} / \NA$ & $\si{\mol \per\second}$ & \cite{Goers2018, Erokhova2016} \\ \hline
        $\ghl$ & $\num{3e-6} \cdot 4 \pi \rTX^2$ & $\si{\cubic \meter \per \second}$ & \cite{tomDieck2025} \\ \hline
        $\gsl$ & $\num{1e-12} \cdot 4 \pi \rTX^2$ & $\si{\cubic \meter \per \second}$ & \cite{Maurer2001} \\ \hline
        $\ghp$ & $\num{700} / \NA$ & $\si{\mol \per \second}$ & \cite{Goers2018,Feldbauer2009}\\ \hline
    \end{tabular}}
    \label{tab:params}
    \vspace{-0.5cm}
\end{table}

\noindent In this section, we analyze the proposed modulation schemes. First, we compare the different methods for calculating $\Tload$ and $\Tdeload$. Then, we analyze the received signal for the modulation schemes introduced in Section~\ref{sec:modulation}. Finally, we compare the performance of the different modulation schemes in terms of \ac{BER}. An overview of the parameter values used is given in \cref{tab:params}. $\NA$ denotes the Avogadro constant.

\subsection{Calculation of $\Tload$ and $\Tdeload$}

\begin{figure}
    \centering
    \includegraphics[width=0.48\textwidth]{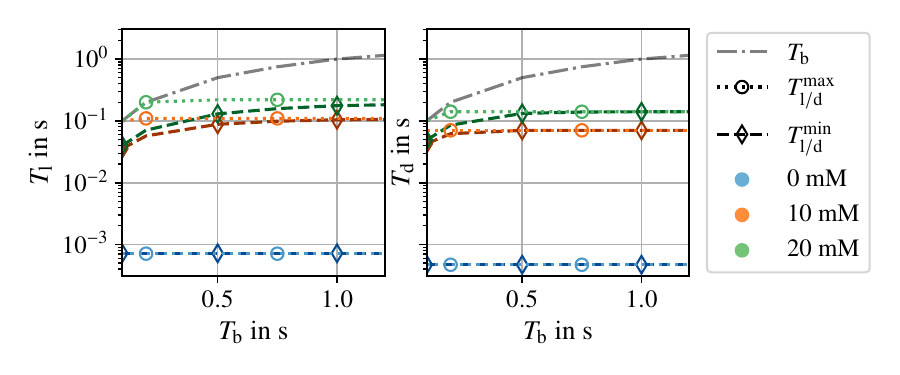}
    \vspace{-1.5em}
    \caption{$\Tload$ (left) and $\Tdeload$ (right) for different $\Tbit$, buffer molarities, and methods of computation. Dashed and dotted curves correspond to $\TloadMin$ and $\TloadMax$, respectively.}
    \vspace{-1.5em}
    \label{fig:tload}
\end{figure}

\noindent Since the calculation of $\Tload$ and $\Tdeload$ is crucial for both \ac{MAM} and \ac{MEM}, we first analyze the calculation of the lower and upper bounds. \Cref{fig:tload} shows the resulting values for varying buffer molarities $\CZERO$ and bit durations for the two considered computation methods. Note that a higher $\CZERO$ corresponds to a higher attenuation of the $\Hplus$ influx over the vesicle membrane. The gray dash-dotted curves represent the bit interval $\Tbit$, which serves as maximum value for $\Tload$ and $\Tdeload$, as \ac{MAM} and \ac{MEM} are not feasible for $\Tload,\Tdeload \geq \Tbit$.
When comparing the $\Tload$ calculation methods, it becomes clear that $\TloadMax$ remains constant for different $\Tbit$ values because it is based on the fixed initial concentration $\ChinS = \ChinZERO$. A comparison of the two subfigures of \cref{fig:tload} further shows that the deloading times are generally shorter than the loading times. This can be attributed to the lower buffer flux attenuation at the respective $\Chin{t}$ values and the absence of $\Hplus$ influx from the energizing module during deloading. 
For both $\TloadMin$ and $\TdeloadMin$, the lower bounds approach the upper bounds as $\Tbit$ increases. This behavior is expected since longer bit durations reduce the difference between $\ChinS$ and $\ChinZERO$ (for $\TloadMin$) and between $\ChinS$ and $\ChinEq$ (for $\TdeloadMin$).
Because of their adaptive behavior, in all subsequent simulations, we use $\TloadMin$ and $\TdeloadMin$ as the loading and deloading times for \ac{MAM} and \ac{MEM}, respectively.

\subsection{Memory Effects in Realizable \ac{MC} Systems}
\begin{figure*}
    \centering
    \includegraphics[width=\linewidth]{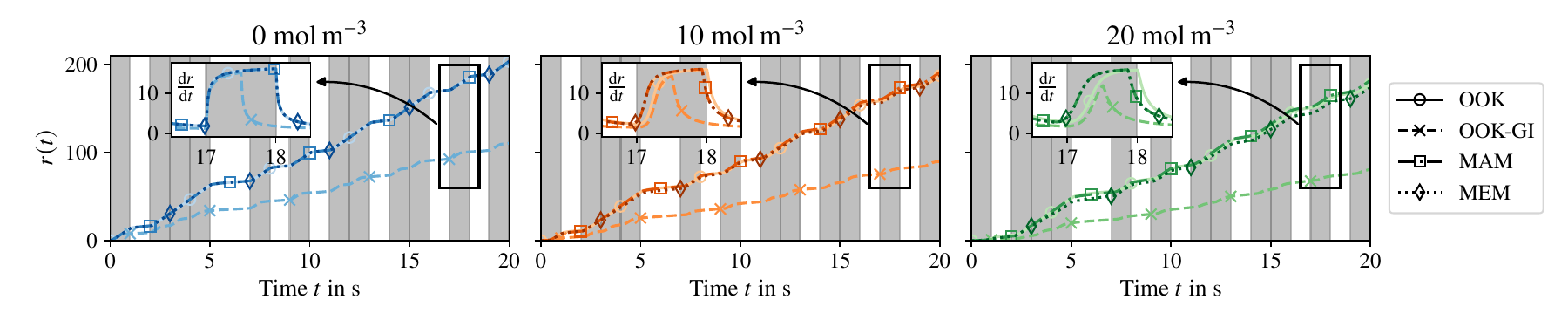}
    \vspace{-2.5em}
    \caption{Expected received signal $r(t)$ for a random bit sequence, varying buffer molarities, and modulation schemes. Gray-shaded areas indicate bit ``1'' intervals and inlet plots show the derivative of $r(t)$ with respect to time, i.e., $\mathrm{d}r(t)/\mathrm{d}t$.}
    \vspace{-1.8em}
    \label{fig:result_memory}
\end{figure*}

\noindent Next, we analyze the expected received signal for the different modulation schemes introduced in \cref{sec:modulation}. \Cref{fig:result_memory} shows the expected received signal $r(t)$ for the different modulation schemes for three different buffer molarities corresponding to the three subfigures. We observe that during bit ``1'' intervals (indicated by gray shading), $r(t)$ shows a higher slope, i.e., an increased rate of received \acp{SM} for all modulation schemes and buffer molarities. For all schemes, the slope during bit ``1'' intervals is largest for the unbuffered scenario (left subplot in \cref{fig:result_memory}) and decreases for increasing buffer molarities $\CZERO$ (middle and right subplot in \cref{fig:result_memory}), which can be explained by the delay of \ac{SM} release caused by the buffer. For \ac{OOK-GI}, we observe that the number of released \acp{SM} is much lower than for the other modulation schemes, as $\TGuard = 0.5 \Tbit$. Therefore, \ac{OOK-GI} is less resource-intensive than the other schemes but due to the low \ac{SM} count also more prone to detection errors. For $\CZERO = \SI{20}{\mol\per\cubic\meter}$(right-hand subfigure of \cref{fig:result_memory}), we can clearly observe the positive effects of \ac{MAM} and \ac{MEM} during bit ``1'' intervals as the slopes of the curves change more rapidly after the beginning of the bit interval (in comparison to \ac{OOK} and \ac{OOK-GI}). This is visualized in the inlets that show the time derivative of $r(t)$, $\frac{\mathrm{d}r(t)}{\mathrm{d}t}$, and where the curves for \ac{MAM} and \ac{MEM} increase sooner after the beginning of the bit ``1'' interval, corresponding to a successful reduction of delay. To analyze the performance of the different modulation schemes in detail, we will next compare their \acp{BER}.

\subsection{Performance of Different Modulation Schemes}

\begin{figure}
    \centering
    \vspace{-0.05cm}
    \includegraphics[width=0.48\textwidth]{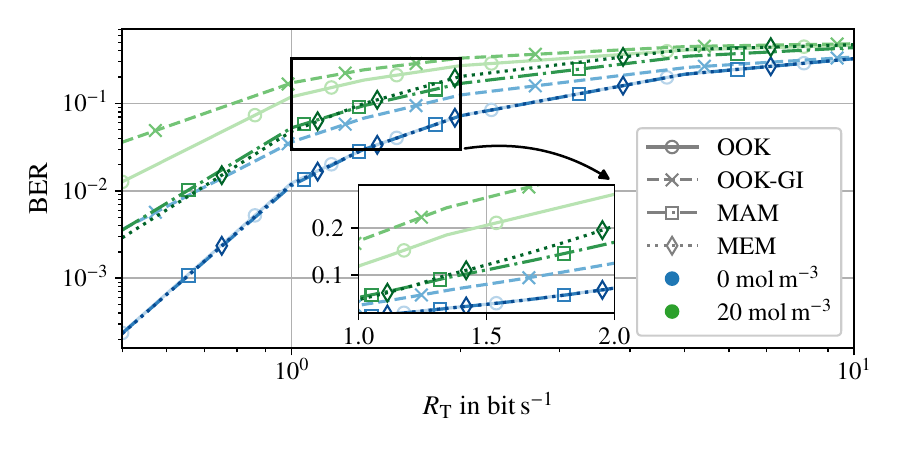}
    \vspace{-1.75em}
    \caption{\Ac{BER} for varying transmission rates $\Rate$ for \ac{OOK}, \ac{OOK-GI}, \ac{MAM}, and \ac{MEM}. Differently colored curves correspond to different buffer molarities $\CZERO = \{0, 20\}$ \si{\mol\per\cubic\meter}.}
    \label{fig:ber}
    \vspace{-2em}
\end{figure}

\noindent \Cref{fig:ber} shows the empirical \ac{BER} for different data rates $\Rate = 1 / \Tbit$ and modulation schemes, where \ac{OOK} and \ac{OOK-GI} serve as baselines, in an unbuffered and a buffered scenario. The empirical \acp{BER} were obtained by adopting a single-sample threshold detection scheme \cite{Huang2022} utilizing $\num{50000}$ Monte Carlo--simulations (to account for propagation noise) simulating $20$ random bit sequences of length $L = 200$. To limit performance-deteriorating effects stemming from the detection itself, we used line search to find the optimal detection threshold for each setting (i.e., each specific bit sequence, buffer molarity, and modulation scheme). The detection sample is taken at the end of each bit interval, i.e., the sampling time for the $i$-th bit is $t_i = (i + 1) \Tbit$. The detection sample of the $i$-th bit, $D_i$, is given by $R(t_i) - R(t_{i-1})$, where $R(t)$ is a Binomial random variable with mean $r(t)$, to obtain the number of \acp{SM} received by the \ac{FAR} during the $i$-th bit interval. It can be observed that the \ac{BER} increases for increasing $\Rate$, i.e., shorter bit intervals. Additionally, the \ac{BER} is lower for lower buffer molarities. Evidently, due to the \ac{GI}, \ac{OOK-GI} performs much worse than all other modulation schemes in all scenarios.
In the unbuffered scenario the performance difference between \ac{OOK}, \ac{MAM}, and \ac{MEM} is marginal. However, the advantages of the proposed modulation schemes become apparent in the buffered scenario, which is relevant for future \ac{MC} systems, where buffers are likely to be present. \ac{MAM} outperforms both \ac{MEM} and \ac{OOK} in the buffered scenario at high $\Rate$. In this case, the default (de)loading intervals at the end of each bit in \ac{MEM} reduce the differentiability between bit ``1''s and ``0''s due to the soft activation of the symporters. However, at lower $\Rate$, \ac{MAM} and \ac{MEM} are comparable in performance. As \ac{MEM} can be applied in scenarios where the next transmit symbol is not known, it thus constitutes a valuable alternative to \ac{MAM}. In summary, \cref{fig:ber} shows that in realistic scenarios involving buffering effects, delay between the modulation signal and the release of \acp{SM}, leakage, and imperfect release control, the proposed modulation schemes \ac{MAM} and \ac{MEM} achieve superior performances compared to \ac{OOK} and \ac{OOK-GI}.

\section{Conclusion and Future Work}
\label{sec:conclusion}
\noindent In this work, we tackle the \ac{TX}-induced delay in \ac{SM} release in \ac{MC} systems. The delay may stem from biological building blocks, which are expected to be key components in future \ac{MC} \ac{TX} implementations. We proposed two modulation schemes to mitigate the related undesired and performance-deteriorating effects directly at the \ac{TX} without introducing a loss in data rate. These modulation schemes were then compared with respect to their performance in simulations. The proposed \ac{MAM} and \ac{MEM} schemes improved the performance of the communication link for different data rates. Additionally, they proved effective for different buffer molarities that are expected in realistic \ac{MC} systems. In future work, the possibility of further performance enhancement by suitable encoding schemes for realizable \ac{MC} systems will be considered. Such schemes may be able to further reduce memory effects or replace tailor-made modulation schemes, e.g., when the modulator in an \ac{MC} system is difficult to control. 

\appendix
\label{app}
\subsection{Calculation of $\ChinEq$ for $n = 3$}
\label{app:B}

\noindent The equilibrium $\Hplus$ concentration, $\ChinEq$, is the solution to 

\begin{equation}
    0 = \underbrace{a \ChinEq + b}_{\text{Pumps and leakage}} + \underbrace{x_1 \frac{(\ChinEq)^n}{(\ChinEq)^n + x_2}}_{\text{Symporters}},
    \label{eq:tload1}
\end{equation}

\noindent with auxiliary variables $x_1 = \ghsym \frac{\CsinNot}{\CsinNot + \Km}$ and $x_2 = \KmH^n$. The parameters $a$ and $b$ capture the influence of the proton flux caused by the pumps and leakage as defined in \cite[Eq. (10)]{tomDieck2025}.
The second part of \eqref{eq:tload1} captures the proton flux caused by the symporters $\ihsym{t}$, i.e., \eqref{eq:hill}. Therefore, \eqref{eq:tload1} corresponds to a net flux of 0 of $\Hplus$ between the in- and outside of the vesicle at a given \ac{SM} concentration. As \eqref{eq:tload1} cannot be solved for arbitrary $\Hill$, we derive $\ChinEq$ for the special case of $n = 3$ as this value is used in this paper. Solving \eqref{eq:tload1} at a specific $\CsinNot$ for $\ChinEq$ yields

\allowdisplaybreaks
\begin{align}
    & \ChinEq = -\frac{X_3}{4} \pm \frac{1}{2} \left (
        \frac{X_3^2}{4}
        - X_2
        - \frac{8}{X_1^3 3} \right )^{1/2} \nonumber \\ & \pm \frac{1}{2} \left (
        \frac{X_3^2}{2} + X_2 + \frac{8}{X_1^3 3a} 
        + \frac{X_3^3 + 8 x_2}{4 \sqrt{
                \frac{X_3^2}{4}
                - X_2
                - \frac{8}{X_1^3 3a}}}
    \right )^{1/2}, \nonumber \\
        X_1 = & -27 b (b + x_1)^2 x_2 - 27 a^3 x_2^2 + \nonumber \\ & \sqrt{729 x_2^2 \left(b^3 + 2 b^2 x_1 + b x_1^2 + a^3 x_2\right)^2 - 4 (9 a b x_2 - 3 a x_1 x_2)^3}, \nonumber \\
    X_2 = & \frac{(3 b - x_1) x_2}{8 X_1}, X_3 = \frac{b + x_1}{a}.
    \label{eq:chineq}
\end{align}

\noindent The valid solution to \eqref{eq:chineq} satisfies $\ChinEq \ge 0$ and $\ChinEq \in \mathbb{R}$, where $\mathbb{R}$ denotes the set of real numbers.

\subsection{Gauss-Seidel Method for Computation of $\TloadMin$ and $\TdeloadMin$}
\label{app:A}

\noindent As mentioned in \cref{sec:loadinganddeloading}, the value for $\TloadMin$ (or $\TdeloadMin$) has to be obtained by a batch Gauss-Seidel method. This method iteratively updates the value of $\TloadMin$ (or $\TdeloadMin$) by using estimates of $\ChinS$ and vice versa: New estimates of $\ChinS$ are obtained by using the most recent estimate of $\TloadMin$ (or $\TdeloadMin$). Note that the estimates have to be obtained simultaneously as $\ChinS$ depends on $\TloadMin$ (or $\TdeloadMin$) and $\TloadMin$ (or $\TdeloadMin$) depends on $\ChinS$, as stated in (C3) in \cref{sec:loadinganddeloading}. The computation method is outlined in \cref{al:gs}. In \cref{al:gs}, \textsc{Find }$t(\cdot)$ and \textsc{Find }$ C(\cdot)$ denote functions to find the time it takes to reach $\ChinSwitch$ from $C^{(i - 1)}$ and the next estimate for $C^{(j)}$ from the current estimate of $\TloadMin$ (or $\TdeloadMin$), i.e., $t^{(j)}$, respectively, where $j \in \mathbb{N}$ is the running variable indexing the current iteration. These functions use \ac{FDM} to mitigate (C1) (see \cref{sec:loadinganddeloading}). When $t^{(j)}$ converges, i.e., the change of $t^{(j)}$ in comparison to $t^{(j - 1)}$ lies below the error tolerance $\epsilon$, we have found $\TloadMin$ (or $\TdeloadMin$). 

\begin{algorithm}
  \caption{Computation of $\TloadMin$ (or $\TdeloadMin$)}
  \begin{algorithmic}[1]
    \State \textbf{Input: } Threshold concentration $\ChinThresh$.
    \State Initialize error tolerance $\epsilon$.
    \State Set initial $t^{(0)} = \infty$.
    \State Set initial start concentration $C^{(0)} = \ChinZERO$ (or $\ChinEq$).
    \State Set $j = 1$.
    \Repeat
        \State Determine $t^{(j)} = \textsc{Find } t (C^{(j - 1)}, \ChinThresh)$.
        \State Determine $C^{(j)} = \textsc{Find } C(t^{(j)}, \ChinThresh)$.
        \State Set $j = j + 1$.
    \Until{$|t^{(j - 1)} - t^{(j-2)}| \le \epsilon$}
    \State  \textbf{Output: } $\TloadMin$ (or $\TdeloadMin$) = $t^{(j - 1)}$.
  \end{algorithmic}
  \label{al:gs}
\end{algorithm}

\bibliographystyle{IEEEtran}
\bibliography{literature}
\end{document}